\documentclass[reprint,superscriptaddress,nofootinbib,preprintnumbers,aps,prl]{revtex4-2}

\usepackage{amsmath}
\usepackage{amssymb}
\usepackage{graphicx}
\usepackage{booktabs}
\usepackage{color}
\usepackage{bbold}
\usepackage[colorlinks=true,allcolors={blue!70!black}]{hyperref}
\usepackage{amsmath}
\usepackage{tabularx}
\usepackage{multirow}

\usepackage{xspace}

\newcommand\as{\alpha_{\mathrm{S}}}

\def\to{\rightarrow}

\def\msbar{{\overline {\rm MS}}}

\def\rcut{r_{\rm cut}}

\def\ttbH{\ensuremath{t {\bar t}H}\xspace}

\def\muF{{\mu_{\rm F}}}
\def\muR{{\mu_{\rm R}}}
\def\muIR{{\mu_{\rm IR}}}

\usepackage{scalefnt,pstricks}
\usepackage{cancel}

\newcommand\Matrix{{\sc Matrix}\xspace}

\newcommand\OpenLoops{{\sc OpenLoops}\xspace}
\newcommand\Recola{{\sc Recola}\xspace}

\usepackage{array}
\newcolumntype{L}[1]{>{\raggedright\let\newline\\\arraybackslash\hspace{0pt}}m{#1}}
\newcolumntype{C}[1]{>{\centering\let\newline\\\arraybackslash\hspace{0pt}}m{#1}}
\newcolumntype{R}[1]{>{\raggedleft\let\newline\\\arraybackslash\hspace{0pt}}m{#1}}

\usepackage[font=scriptsize, labelfont=bf]{caption}

\usepackage{lipsum}

\begin{document}

\renewcommand{\thefootnote}{\fnsymbol{footnote}}

\title{Higgs boson production in association with a \\ top--antitop quark pair 
  in next-to-next-to leading order QCD}

\author{Stefano Catani}
\email{catani@fi.infn.it}
\affiliation{INFN, Sezione di Firenze and Dipartimento di Fisica e Astronomia, Universit\`a degli Studi di Firenze, 50019 Sesto Fiorentino, Firenze, Italy}
\author{Simone Devoto}
\email{simone.devoto@unimi.it}
\affiliation{Dipartimento di Fisica, Universit\`{a} degli Studi di Milano and INFN, Sezione di Milano, 20133 Milano, Italy}
\author{Massimiliano Grazzini}
\email{grazzini@physik.uzh.ch}
\affiliation{Physik Institut, Universit\"at Z\"urich, CH-8057 Z\"urich, Switzerland}
\author{Stefan~Kallweit}
\email{stefan.kallweit@cern.ch}
\affiliation{Dipartimento di Fisica, Universit\`{a} degli Studi di Milano-Bicocca and INFN, Sezione di Milano-Bicocca, I-20126, Milan, Italy}
\author{Javier Mazzitelli}
\email{javier.mazzitelli@psi.ch}
\affiliation{Max-Planck-Institut f\"ur Physik, 80805 M\"unchen, Germany}
\affiliation{Paul Scherrer Institut, 5232 Villigen PSI, Switzerland}
\author{Chiara Savoini}
\email{csavoi@physik.uzh.ch}
\affiliation{Physik Institut, Universit\"at Z\"urich, CH-8057 Z\"urich, Switzerland}
%\author{Stefano Catani${}^{(a)}$, Simone Devoto${}^{(b)}$, Massimiliano Grazzini${}^{(c)}$, Stefan Kallweit${}^{(d)}$, Javier Mazzitelli${}^{(e,f)}$ and Chiara Savoini${}^{(c)}$\vspace{1em}}
%\affiliation{${}^{(a)}$INFN, Sezione di Firenze and Dipartimento di Fisica e Astronomia, Universit\`a degli Studi di Firenze, 50019 Sesto Fiorentino, Firenze, Italy}
%\affiliation{${}^{(b)}$Dipartimento di Fisica, Universit\`{a} degli Studi di Milano and INFN, Sezione di Milano, 20133 Milano, Italy}
%\affiliation{${}^{(c)}$Physik Institut, Universit\"at Z\"urich, 8057 Z\"urich, Switzerland}
%\affiliation{$^{(d)}$Dipartimento di Fisica, Universit\`{a} degli Studi di Milano-Bicocca and INFN, Sezione di Milano-Bicocca, 20126 Milano, Italy}
%\affiliation{${}^{(e)}$Max-Planck-Institut f\"ur Physik, 80805 M\"unchen, Germany}
%\affiliation{${}^{(f)}$Paul Scherrer Institut, 5232 Villigen PSI, Switzerland}
\begin{abstract}
The associated production of a Higgs boson with a top--antitop quark pair is a crucial process at the LHC since it
allows for a direct measurement of the top-quark Yukawa coupling. We present the computation of the radiative
corrections to this process at the next-to-next-to-leading order (NNLO) in QCD perturbation theory.
This is the very first computation for a \mbox{$2\to 3$} process with massive coloured particles at this perturbative order.
We develop a soft Higgs boson approximation for loop amplitudes, which enables us to reliably quantify the impact of the yet unknown two-loop contribution.
At the centre-of-mass energy \mbox{$\sqrt{s}=13$\,TeV} the NNLO corrections
increase the next-to-leading order result for the total cross section by about $4\%$ and lead to a significant reduction of perturbative uncertainties.
\end{abstract}

\maketitle

\paragraph*{Introduction.} About ten years ago the ATLAS and CMS collaborations announced the discovery of a scalar resonance~\cite {ATLAS:2012yve,CMS:2012qbp} whose properties resembled those expected for the Higgs boson predicted by the Standard Model (SM). By now, the experimental data have significantly sharpened this picture by assembling information from different production and decay channels, and a framework of Higgs boson interactions has emerged
that is fully consistent with the SM hypothesis~\cite{ATLAS:2022vkf,CMS:2022dwd}.

Since the Higgs boson couplings to SM particles are proportional to their masses,
a special role is played by the coupling to the top quark.
The observation of Higgs boson production in association with a top--antitop quark pair was reported by the ATLAS and CMS collaborations in 2018~\cite{Aaboud:2018urx,Sirunyan:2018hoz}.
This production mode allows for a direct measurement of the top-quark Yukawa coupling.
Any deviation from the SM prediction would be a signal of New Physics.
At present ATLAS and CMS measure the signal strength in this channel to an accuracy of ${\cal O}(20\%)$, but at the end of the high-luminosity phase the uncertainties are expected to reach the ${\cal O}(2\%)$ level~\cite{Cepeda:2019klc}.

The first theoretical studies of the hadroproduction of a top--antitop quark pair and a Higgs boson~(\ttbH) in the SM were carried out in Refs.~\cite{Ng:1983jm,Kunszt:1984ri} at leading order~(LO) in QCD perturbation theory, and in Refs.~\cite{Beenakker:2001rj,Beenakker:2002nc,Reina:2001sf,Reina:2001bc,Dawson:2002tg,Dawson:2003zu} at next-to-leading order~(NLO).
NLO electroweak~(EW) corrections were reported in Refs.~\cite{Frixione:2014qaa,Yu:2014cka,Frixione:2015zaa}.
The resummation of soft-gluon contributions close to the partonic threshold was considered in Refs.~\cite{Kulesza:2015vda,Broggio:2015lya,Broggio:2016lfj,Kulesza:2017ukk,Broggio:2019ewu,Ju:2019lwp,Kulesza:2020nfh}.
Full off-shell calculations with decaying top quarks were presented at NLO QCD~\cite{Denner:2015yca} and NLO QCD+EW~\cite{Denner:2016wet}.
The current theoretical uncertainties of the $\ttbH$ cross section are at the ${\cal O}(10\%)$ level~\cite{LHCHiggsCrossSectionWorkingGroup:2016ypw}.
To match the experimental precision expected at the end of the high-luminosity phase of the LHC, next-to-next-to-leading order~(NNLO) predictions in QCD perturbation theory are indispensable.

The NNLO calculation of \ttbH production requires
the evaluation of tree-level contributions with two additional
unresolved partons in the final state,
of one-loop contributions with one unresolved parton, and
of purely virtual contributions.
The required tree-level and one-loop scattering amplitudes can nowadays be evaluated with automated tools.
The two-loop amplitude for \ttbH production is not known and its computation is
at the frontier of current possibilities~\cite{Heinrich:2020ybq,Chen:2022nxt}.

Even with all the required amplitudes available,
their implementation in a complete
NNLO calculation is
a difficult task because of the presence of infrared~(IR)
divergences at intermediate stages of the calculation.
Various methods have been proposed and used to overcome these difficulties at the NNLO level (see Refs.~\cite{Bendavid:2018nar,Amoroso:2020lgh,TorresBobadilla:2020ekr,Heinrich:2020ybq} and references therein).

In this Letter we will use the transverse-momentum~($q_T$) subtraction method~\cite{Catani:2007vq}.
The method uses IR subtraction counterterms that are constructed by considering
the $q_T$ distribution
of the produced final-state system in the limit \mbox{$q_T \to 0$}~\cite{Catani:2013tia,Zhu:2012ts,Li:2013mia,Catani:2014qha}.
Originally developed for the production of a colour singlet, the method has been extended to heavy-quark production and applied to the NNLO computations of top-quark and bottom-quark pair production~\cite{Bonciani:2015sha,Catani:2019iny,Catani:2019hip,Catani:2020kkl}.

The production of a heavy-quark pair accompanied by a colourless particle does not pose any additional conceptual complications in the context of the $q_T$ subtraction formalism.
However, its implementation requires the computation of
appropriate soft-parton contributions.
The results of this computation at NLO and, partly, at NNLO were presented in Ref.~\cite{Catani:2021cbl},
and the evaluation of the NNLO soft terms has been subsequently completed~\cite{Catani:2023tby,inprep}. 
Following the NNLO computation of the off-diagonal partonic channels~\cite{Catani:2021cbl},
in this Letter we present the NNLO result for $\ttbH$ production including all the partonic channels.
The two-loop amplitudes are not yet known, and we evaluate them by developing and using a soft Higgs boson approximation.
As we will show, this approximation allows us to obtain the NNLO corrections with very small residual uncertainties.

\paragraph*{The calculation.}

We consider the process
\begin{equation}
\label{eq:process}
c(p_1)+{\bar c}(p_2)\to t(p_3)+{\bar t}(p_4)+H(k)\,,\qquad c=q,{\bar q},g,
\end{equation}
where the collision of the massless partons of flavours $c$ and $\bar c$ and momenta $p_1$ and $p_2$ produces
a top-antitop quark pair of momenta $p_3$ and $p_4$, and a Higgs boson with momentum $k$.
We denote the pole masses of the top quark and the Higgs boson by $m_t$ and $m_H$, respectively.

The renormalised all-order
scattering amplitude for the process in Eq.~(\ref{eq:process}) is denoted as ${\cal M}(\{p_i\},k)$.
In the limit in which the Higgs boson becomes soft (\mbox{$k\to 0$}),
${\cal M}(\{p_i\},k)$ fulfils the following factorisation formula:
\begin{equation}
\label{eq:fact}
{\cal M}(\{p_i\},k)\simeq F(\as(\muR);\tfrac{m_t}{\muR})\,\frac{m_t}{v}\sum_{i=3,4} \frac{m_t}{p_i \cdot k}\, {\cal M}(\{p_i\})\, ,
\end{equation}
where \mbox{$v=(\sqrt{2}G_F)^{-1/2}=246.22$\,GeV} and ${\cal M}(\{p_i\})$ is the amplitude in which the Higgs boson has been removed.
Details on the derivation of Eq.~(\ref{eq:fact}) and the explicit expression of the perturbative function $F(\as(\muR); m_t/\muR)$ at the second order in the QCD coupling $\as(\muR)$ can be found in the Supplemental Material.
Since the virtual amplitudes ${\cal M}(\{p_i\})$ for the production of a $t{\bar t}$ pair are available up to two-loop order~\cite{Barnreuther:2013qvf}, the factorisation formula in Eq.~(\ref{eq:fact}) can be used to provide an approximation of the virtual $t{\bar t}H$ amplitudes up to the same order.

The cross section for \ttbH production can be written as
$\sigma=\sigma_{\rm LO}+\Delta\sigma_{\rm NLO}+\Delta\sigma_{\rm NNLO}+\dots$,
where $\sigma_{\rm LO}$ is the LO cross section, $\Delta\sigma_{\rm NLO}$ is the NLO QCD correction, $\Delta\sigma_{\rm NNLO}$ is the NNLO QCD contribution, and so forth.

According to the $q_T$ subtraction formalism~\cite{Catani:2007vq} the differential cross section $d\sigma$ can be evaluated as
\begin{equation}
\label{eq:main}
d\sigma={\cal H}\otimes d\sigma_{\rm LO}+\left[d\sigma_{\rm R}-d\sigma_{\rm CT}\right]\, .
\end{equation}
The first term on the r.h.s.\ of Eq.~(\ref{eq:main}) corresponds to the \mbox{$q_T=0$} contribution.
It is obtained through a convolution, with respect to the longitudinal-momentum fractions $z_1$ and $z_2$ of the colliding partons,
of the perturbative function ${\cal H}$ with the LO cross section $d\sigma_{\rm LO}$.
The real contribution $d\sigma_{\rm R}$ is obtained by evaluating the cross section to produce the \ttbH system accompanied
by additional QCD radiation.
When $d\sigma$ is computed at NNLO, $d\sigma_{\rm R}$ contains NLO-type singularities and can be evaluated for example by using the dipole subtraction formalism~\cite{Catani:1996jh,Catani:1996vz,Catani:2002hc}.
The role of the counterterm $d\sigma_{\rm CT}$ is to cancel the singular behaviour of $d\sigma_{\rm R}$ in the limit \mbox{$q_T\to 0$}, making the square-bracket term in Eq.~(\ref{eq:main}) finite. The explicit form of $d\sigma_{\rm CT}$ is known up to NNLO and obtained from the perturbative expansion of the $q_T$ resummation formula for \ttbH production~\cite{Catani:2014qha,Catani:2021cbl}.

Our computation is implemented within the \Matrix framework~\cite{Grazzini:2017mhc}, suitably extended to \ttbH production, along the lines of what has been done for heavy-quark production~\cite{Catani:2019iny,Catani:2019hip,Catani:2020kkl}.
The required tree-level and one-loop amplitudes are obtained with \OpenLoops~\cite{Cascioli:2011va, Buccioni:2017yxi,Buccioni:2019sur}.
To  numerically evaluate the contribution in the square bracket of Eq.~(\ref{eq:main}), a technical cut-off $\rcut$ is introduced on the variable $q_T / M$, where $M$ is the invariant mass of the $t\bar t H$ system. The final result, which corresponds to the limit \mbox{$\rcut\to 0$}, is extracted by computing the cross section at fixed values of $\rcut$ and performing the \mbox{$\rcut\to 0$} extrapolation.
More details can be found in Refs.~\cite{Grazzini:2017mhc,Catani:2021cbl}.

We come back to the first term on the r.h.s.\ of Eq.~(\ref{eq:main}).
The function ${\cal H}$ can be decomposed as
\begin{equation}
\label{eq:deco}
{\cal H}=H\delta(1-z_1)\delta(1-z_2)+\delta{\cal H}\, ,
\end{equation}
where the hard coefficient $H$ contains purely virtual contributions and flavour indices are understood. More precisely, we define
\begin{equation}
H(\as(\muR);\tfrac{M}{\muIR})=1+\sum_{n=1}^\infty \left(\tfrac{\as(\muR)}{2\pi}\right)^n H^{(n)}(\tfrac{M}{\muIR})
\end{equation}
with
\begin{equation}
\label{eq:Hn}
H^{(n)}=\left.\frac{2{\rm Re}\left({\cal M}^{(n)}_{\rm fin}(\muIR,\muR){\cal M}^{(0)*}\right)}{|{\cal M}^{(0)}|^2}\right\vert_{\muR=M}\, .
\end{equation}
Here ${\cal M}^{(0)}$ is the Born level amplitude for the \mbox{$c{\bar c}\to t{\bar t}H$} process, while ${\cal M}^{(n)}_{\rm fin}$ are the perturbative coefficients of the finite part ${\cal M}_{\rm fin}(\muIR)$ of the renormalised virtual amplitude 
after subtraction of IR singularities at the scale $\muIR$.
The IR-finite amplitude ${\cal M}_{\rm fin}$
is obtained from the all-order renormalised virtual amplitude ${\cal M}$ as 
\mbox{$|{\cal M}_{\rm fin}(\muIR)\rangle={\mathbf Z}^{-1}(\muIR)| {\cal M}\rangle$},
where ${\mathbf Z}(\muIR)$ is the multiplicative factor that removes the IR $\epsilon$ poles of the multiloop amplitude~\cite{Ferroglia:2009ii}.
In the case of \mbox{$n=2$} the definition of $H^{(n)}$ in Eq.~(\ref{eq:Hn}) allows us to isolate the only unknown contribution to the NNLO cross section in Eq.~(\ref{eq:main}). Indeed at NNLO the square bracket in Eq.~(\ref{eq:main}) is computable for all the partonic channels, and the other contributions, embodied in the function $\delta{\cal H}^{(n)}$ on the r.h.s.\
of Eq.~(\ref{eq:deco}), are completely known.
In particular, at NNLO $\delta{\cal H}$ also includes the one-loop squared contribution and the soft-parton contributions~\cite{inprep}.
We point out that, if all the perturbative ingredients are available, the dependence on $\muIR$ exactly cancels out between $H$ and $\delta{\cal H}$ in Eq.~(\ref{eq:deco}). 

In the following we will use the factorisation formula in Eq.~(\ref{eq:fact}) to construct approximations
of the coefficients $H^{(1)}$ and $H^{(2)}$.
To do so, we need to introduce a prescription that, from an event containing a $t{\bar t}$ pair and a Higgs boson, defines a corresponding event in which the Higgs boson is removed. We will use the $q_T$ recoil prescription~\cite{Catani:2015vma}, where the momenta of the top and the antitop quark are left unchanged and the transverse momentum of the Higgs boson is equally reabsorbed by the initial-state partons. This prescription is used to evaluate the $t{\bar t}$ amplitudes on the r.h.s.\ of the factorisation formula in Eq.~(\ref{eq:fact}).
We also need to define the subtraction scale $\muIR$. In the evaluation of the $H^{(1)}$ and $H^{(2)}$ contributions the subtraction scale $\muIR$ is set to the virtuality of the \ttbH system.
In the IR subtracted $t{\bar t}$ amplitudes required to evaluate the factorisation formula, we will set $\muIR$ to the virtuality of the $t{\bar t}$ pair. At tree-level and one-loop order the $t{\bar t}$ amplitudes are obtained with \OpenLoops~\cite{Cascioli:2011va, Buccioni:2017yxi,Buccioni:2019sur}, while at two-loop order we use the results of Ref.~\cite{Barnreuther:2013qvf}.

Our numerical results are obtained for proton--proton collisions at the centre-of-mass energies between \mbox{$\sqrt{s}=8\,\mathrm{TeV}$} and \mbox{$\sqrt{s}=100\,\mathrm{TeV}$}.
We use the NNLO NNPDF31~\cite{Ball:2017nwa} parton distribution functions~(PDFs)
throughout, with the QCD running coupling $\as$ evaluated
at 3-loop order. The pole mass of the top quark is \mbox{$m_t=173.3$\,GeV}, the Higgs boson mass \mbox{$m_H=125$\,GeV}, and the Fermi constant \mbox{$G_F = 1.16639\times 10^{-5}$\,GeV$^{-2}$}.
The central values of the renormalisation and factorisation scales, $\muR$ and $\muF$, are fixed at \mbox{$\muR=\muF=(2m_t+m_H)/2$}.

To validate our soft Higgs boson approximation, we first study its quality at LO.
We compare the LO cross sections $\sigma_{\rm LO}$
in the $gg$ and $q{\bar q}$ partonic channels
with the corresponding approximated results from the soft factorisation formula in Eq.~(\ref{eq:fact}). In the $gg$ channel the result obtained in the soft approximation is a factor of 2.3\,(2) larger than the exact result at \mbox{$\sqrt{s}=13\,(100)$\,TeV}. The situation is better for the $q{\bar q}$ channel, where the soft approximation is only a factor 1.11\,(1.06) larger than the exact result. Despite the fact that the physical Higgs boson is far from the kinematical region for which Eq.~(\ref{eq:fact}) is derived, the soft approximation gives the right order of magnitude of the LO cross section.

We now move to NLO
and compute the contribution $\Delta\sigma_{\rm NLO,H}$ of the coefficient $H^{(1)}$ to the NLO correction
and its soft approximation, $\Delta\sigma_{\rm NLO,H}|_{\rm soft}$. Note that in the soft approximation both numerator and denominator of Eq.~(\ref{eq:Hn}) are evaluated in the soft limit, i.e.\ we define
\begin{equation}
\label{eq:Hns}
H^{(n)}|_{\rm soft}=\left.\frac{2{\rm Re}\left({\cal M}^{(n)}_{\rm fin}(\muIR,\muR){\cal M}^{(0)*}\right)_{\rm soft}}{|{\cal M}^{(0)}|_{\rm soft}^2}\right\vert_{\muR={\tilde M}}\, ,
\end{equation}
where ${\tilde M}$ is the virtuality of the $t{\bar t}$ pair.
By using this approximation we are effectively reweighting the exact LO cross section appearing in the first term in Eq.~(\ref{eq:main}).
This is expected to be a better approximation than simply computing the numerator in the soft limit, since the effect of the soft approximation largely cancels out in the ratio. 

The results are shown in Table~\ref{tab:soft} for both the $gg$ and $q{\bar q}$ channels: In the $gg$ channel the $\Delta\sigma_{\rm NLO,H}$ contribution is underestimated in the approximation by just about $30\%$ on the inclusive level at both collider energies.  We find that this deviation depends only mildly on kinematic variables, which suggests that the good agreement is not due to some accidental cancellation between different phase space regions.
The approximation works even better for the $q{\bar q}$ channel, where the exact result is underestimated by only $5\%$.
The observed deviation for $\Delta\sigma_{\rm NLO,H}$ can be used as an estimate of the uncertainty in our approximation of $\Delta\sigma_{\rm NNLO,H}$.

\renewcommand\arraystretch{1.5}
\begin{table}
\begin{center}
\resizebox{0.47\textwidth}{!}{
\begin{tabular}{|c|ll|ll|}\hline
& \multicolumn{2}{c|}{$\sqrt{s}=13\,\mathrm{TeV}$} & \multicolumn{2}{c|}{$\sqrt{s}=100\,\mathrm{TeV}$}\\
\hline
$\sigma$ [fb] & \multicolumn{1}{c}{$gg$} & \multicolumn{1}{c|}{$q{\bar q}$} & \multicolumn{1}{c}{$gg$} & \multicolumn{1}{c|}{$q{\bar q}$}\\
\hline
$\sigma_{\rm LO}$ & $\phantom{-}261.58$ & $129.47$ & $\phantom{-}23055$ & $2323.7$\\
\hline
$\Delta\sigma_{\rm NLO,H}$ & $\phantom{-0}88.62$  & $\phantom{00}7.826$ & $\phantom{-0}8205$ & $\phantom{0}217.0$\\
$\Delta\sigma_{\rm NLO,H}|_{\rm soft}$ & $\phantom{-0}61.98$ & $\phantom{00}7.413$ & $\phantom{-0}5612$ & $\phantom{0}206.0$\\
\hline
$\Delta\sigma_{\rm NNLO,H}|_{\rm soft}$ & $\phantom{00}{-}2.980(3)$ & $\phantom{00}2.622(0)$ & $\phantom{00}{-}239.4(4)$ & $\phantom{00}65.45(1)$\\
\hline
\end{tabular}
}
\end{center}
\caption{\label{tab:soft}Hard contribution to the NLO and NNLO cross sections in the soft approximation.
Results are shown for the $gg$ and $q{\bar q}$ partonic channels for $\sqrt{s}=13$ TeV and $\sqrt{s}=100$ TeV. Exact results at LO and NLO are shown for comparison.}
\end{table}

Our results for $\Delta\sigma_{\rm NNLO,H}|_{\rm soft}$, which is the contribution of $H^{(2)}$ to the NNLO cross section in the soft approximation, are reported
in the last row of Table~\ref{tab:soft}. In the $gg$ ($q{\bar q}$) channel $\Delta\sigma_{\rm NNLO,H}|_{\rm soft}$ is about $1\%$ ($2\%-3\%$) of the LO cross section.
Therefore, we can anticipate that at NNLO the uncertainties due to the soft approximation will be rather small.

We have repeated our calculation by using other variants of the recoil prescription of Ref.~\cite{Catani:2015vma}, for example by reabsorbing
the transverse momentum of the Higgs boson entirely into one of the initial-state momenta: we find that the results are very close to those
obtained with the symmetric prescription, leading to a negligible uncertainty compared to that derived below. We have also varied the infrared subtraction scale $\muIR$ at which the soft approximation is applied, by repeating the computation for \mbox{$\muIR=M/2$} and \mbox{$\muIR=2M$} while adding the exact evolution terms from $M/2$ and $2M$ to $M$, respectively. The NNLO contribution $\Delta\sigma_{\rm NNLO,H}|_{\rm soft}$ changes by ${}^{+164\%}_{-25\%}$ (${}^{+142\%}_{-20\%}$) in the $gg$ channel and by ${}^{+4\%}_{-0\%}$ (${}^{+3\%}_{-0\%}$) in the $q{\bar q}$ channel at \mbox{$\sqrt{s}=$ 13(100)\,TeV}.

To provide a conservative estimate of the uncertainty, we start from the NLO results.
As discussed above, at NLO the soft approximation underestimates $\Delta\sigma_{\rm NLO,H}$ by $30\%$ in the $gg$ channel and by $5\%$ in the $q{\bar q}$ channel.
Therefore, the uncertainty on $\Delta\sigma_{\rm NNLO,H}|_{\rm soft}$ cannot be expected to be smaller than these values.
We multiply this uncertainty by a tolerance factor that is chosen to be 3 for both the $gg$ and the $q{\bar q}$ channels, taking into account the overall quality of the approximation and the effect of the $\muIR$ variations discussed above.
To obtain the final uncertainty on
the full NNLO cross section, we linearly combine the ensuing uncertainties from the $gg$ and $q{\bar q}$ channels.
As we will see, the overall uncertainty on the NNLO cross section estimated in this way is still significantly smaller than the residual perturbative uncertainties.

\paragraph*{Results.}
\label{sec:resu}

We are now ready to present our results for the inclusive $\ttbH$ cross section.
In Table~\ref{tab:res} we report LO, NLO and NNLO cross sections.
The scale uncertainties are obtained
through the customary procedure of independently varying the renormalisation ($\muR$) and factorisation ($\muF$) scales by a factor of 2 around their central value with the constraint \mbox{$0.5 \leq \muR/\muF \leq 2$}.
Since, as can be seen from Table~\ref{tab:res}, such scale uncertainties are highly asymmetric, especially at NNLO, in the following we will conservatively consider their symmetrised version as our estimate of perturbative uncertainty. More precisely, we take the maximum among the upward and downward variations, assign it symmetrically
and leave the nominal prediction unchanged.

\begin{table}[t]
\begin{center}    
\begin{tabular}{|c|l|l|}\hline
$\sigma$ [pb] & \multicolumn{1}{c|}{$\sqrt{s}=13\,\mathrm{TeV}$} & \multicolumn{1}{c|}{$\sqrt{s}=100\,\mathrm{TeV}$}\\
\hline
$\sigma_{\rm LO}$ & $0.3910\,^{+31.3\%}_{-22.2\%}$ & $25.38\,^{+21.1\%}_{-16.0\%}$\\
$\sigma_{\rm NLO}$ & $0.4875\,^{+5.6\%}_{-9.1\%}$ & $36.43\,^{+9.4\%}_{-8.7\%}$\\
$\sigma_{\rm NNLO}$ & $0.5070\,(31)^{+0.9\%}_{-3.0\%}$ & $37.20(25)\,^{+0.1\%}_{-2.2\%}$\\
\hline
\end{tabular}
\end{center}
\caption{\label{tab:res}LO, NLO and NNLO cross sections at \mbox{$\sqrt{s}=13$\,TeV} and \mbox{$\sqrt{s}=100$\,TeV}. The errors stated in brackets at NNLO combine numerical errors with the uncertainty due to the soft Higgs boson approximation.}
\end{table}

The errors stated in brackets at NNLO are obtained by combining the uncertainty from the soft Higgs boson approximation, estimated as discussed above, with the (much smaller) systematic uncertainty from the subtraction procedure.
Comparing NLO and LO results we see that NLO corrections increase the LO result by $25\%$ at \mbox{$\sqrt{s}=13$\,TeV} and by $44\%$ at \mbox{$\sqrt{s}=100$\,TeV}.
The impact of NNLO corrections is much smaller: they increase the NLO result by $4\%$ at \mbox{$\sqrt{s}=13$\,TeV} and by $2\%$ at \mbox{$\sqrt{s}=100$\,TeV}.
The NNLO contribution of the off-diagonal channels~\cite{Catani:2021cbl} is below the permille level at \mbox{$\sqrt{s}=13$\,TeV}, while it amounts to about half of the computed correction at \mbox{$\sqrt{s}=100$\,TeV}.
Perturbative uncertainties are reduced down to the few-percent level. The uncertainty from the soft Higgs boson approximation amounts to about $\pm 0.6\%$ at both values of $\sqrt{s}$. We point out that this uncertainty, although not negligible, is still significantly smaller than the remaining perturbative uncertainties.

\begin{figure}[t]
\begin{center}
\includegraphics[width=0.4\textwidth]{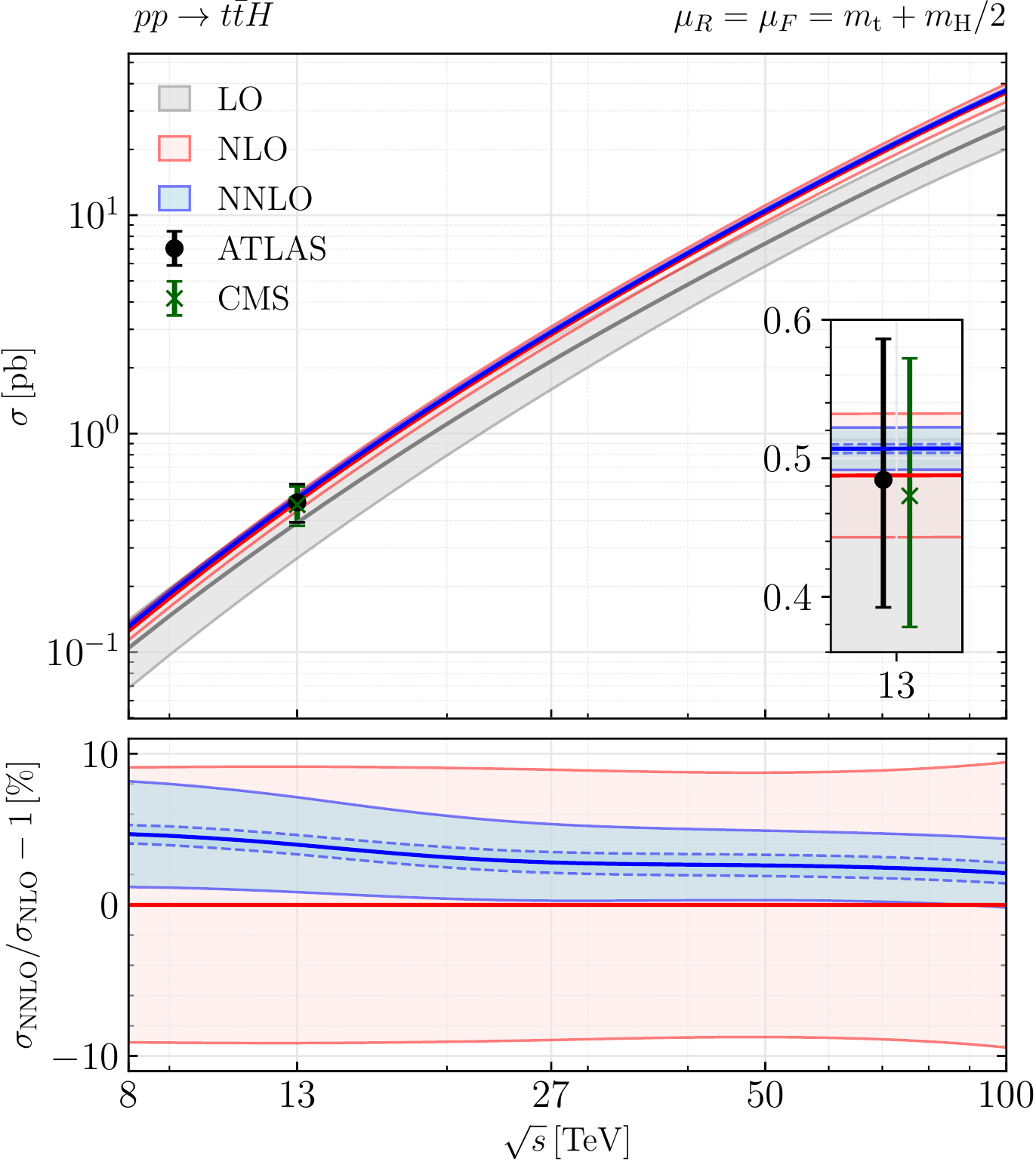}
\end{center}
\caption[]{\label{fig:XS}{LO, NLO and NNLO cross sections with their perturbative uncertainties as functions of the centre-of-mass energy.
The experimental results from ATLAS~\cite{ATLAS:2022vkf} and CMS~\cite{CMS:2022dwd} at \mbox{$\sqrt{s}=13$\,TeV} are also shown.
The lower panel illustrates the impact of NNLO corrections with respect to the NLO result. The inner NNLO band denotes the uncertainty from the soft approximation combined with the systematic uncertainty from the subtraction procedure.}}
\end{figure}
In Fig.~\ref{fig:XS} we show the LO, NLO and NNLO cross sections and their perturbative uncertainties as functions of the centre-of-mass energy $\sqrt{s}$.
The lower panel illustrates the relative impact of the NNLO corrections with respect to the NLO result.
The inner NNLO band denotes the combination of the uncertainty from the soft approximation with the systematic uncertainty from the subtraction procedure.
We see that NNLO corrections range from about $+4\%$ at low $\sqrt{s}$ to about $+2\%$ at \mbox{$\sqrt{s}=100$\,TeV}. The perturbative uncertainty is reduced from $\pm 9\%$ at NLO in the entire range of $\sqrt{s}$ to $\pm 3\%$ ($\pm 2\%$) at \mbox{$\sqrt{s}=8$\,TeV} (\mbox{$\sqrt{s}=100$\,TeV}). We observe that the NNLO band is fully contained within the NLO band.
The experimental results by ATLAS (Fig.~04a in the auxiliary material of Ref.~\cite{ATLAS:2022vkf}) and CMS~\cite{CMS:2022dwd} at \mbox{$\sqrt{s}=13$\,TeV} are also shown for reference in Fig.~\ref{fig:XS}.
We point out that for a sensible comparison with experimental data NLO EW corrections should be considered as well.
At \mbox{$\sqrt{s}=13$\,TeV}, NLO EW corrections increase the cross section by $1.7\%$ with respect to the NLO result~\cite{LHCHiggsCrossSectionWorkingGroup:2016ypw}.

\paragraph*{Summary.}
\label{sec:summa}
The associated production of a Higgs boson with a top--antitop quark pair is a crucial process at hadron colliders since it allows for a direct measurement of the top-quark Yukawa coupling.
In this Letter we have presented first NNLO QCD results for the $t{\bar t}H$ cross section in proton collisions.
The calculation is complete except for the finite part of the two-loop virtual amplitude that is computed by using a soft Higgs boson approximation.
Such approximation is constructed by applying a soft factorisation formula that is presented here, for the first time, up to NNLO in QCD perturbation theory.
This formula will offer strong checks of future exact computations of two-loop amplitudes for processes in which a Higgs boson is produced in association with heavy quarks.
Since the quantitative impact of the genuine two-loop contribution in our computation is relatively small, our approximation allows us to control the NNLO $\ttbH$ cross section to better than $1\%$.
The NNLO corrections are moderate, and range from about $+4\%$ at \mbox{$\sqrt{s}=13$\,TeV} to $+2\%$ at \mbox{$\sqrt{s}=100$\,TeV}, while QCD perturbative uncertainties are reduced to the few-percent level.
When combined with NLO EW corrections, our calculation allows us to obtain the most advanced perturbative prediction to date for the $t{\bar t}H$ cross section.

\vskip 0.5cm
\noindent {\bf Acknowledgements}    

\noindent We would like to thank Thomas Gehrmann and Gudrun Heinrich for helpful discussions and comments on the manuscript.
We are most grateful to Jonas Lindert and the \OpenLoops\ collaboration
for their continuous assistance during the course of this project.
We thank the organisers of ``Loops and Legs in Quantum Field Theory'' (Ettal, Germany, April 2022), where this work was initiated.
This work is supported in part by the Swiss National Science Foundation (SNF) under contract 200020$\_$188464.
The work of SD is supported by the Italian Ministero della Universit\`{a} e della Ricerca (grant PRIN201719AVICI 01).
The work of SK is supported by the ERC Starting Grant 714788 REINVENT.

%\onecolumngrid
%\newpage
\appendix

\section*{Supplemental material}

In this Supplemental Material we provide more details on the derivation of the factorization formula in Eq.~(2).
The perturbative function $F(\as(\muR); m_t/\muR)$
can be extracted from the soft limit of the scalar form factor
of the heavy quark. Up to ${\cal O}(\as^2)$ it reads~\cite{Bernreuther:2005gw,Ablinger:2017hst}
\begin{widetext}
\begin{align}
\label{eq:F}
F(\as(\muR);m_t/\muR)=&\;1+\frac{\as(\muR)}{2\pi} \left(-3\,C_F\right)\nonumber\\
&\hspace*{-7em}+\left(\frac{\as(\muR)}{2\pi}\right)^2\left(\frac{33}{4}\,C_F^2-\frac{185}{12} \,C_F C_A+\frac{13}{6} \,C_F (n_L+1)-6\,C_F\beta_0\ln\frac{\muR^2}{m_t^2}\right)+{\cal O}(\as^3)\, ,
\end{align}
\end{widetext}
where $C_F$ and $C_A$ are the QCD colour factors, $n_L$ is the number of massless flavours,
$\beta_0=\beta^{(n_L)}_0=\frac{11}{12}C_A-\frac{1}{6}n_L$
is the first coefficient of the QCD beta function and $\as(\muR)=\as^{(n_L)}(\muR)$ is the QCD running coupling with $n_L$ active flavours at the renormalisation scale $\muR$\footnote{Note that the result of~\cite{Ablinger:2017hst} is obtained in a scheme with $n_L+1$ active flavours.
To the purpose of our calculation, it is straightforward to check that the
decoupling transformation from $\as^{(n_L+1)}$ to $\as^{(n_L)}$ simply amounts to replacing $\beta_0^{(n_L+1)}$ with $\beta_0^{(n_L)}$ in Eq.~(8).}.

The factorisation formula in Eq.~(2) can be derived by using the eikonal approximation and following the strategy
used for soft-gluon factorisation~\cite{Catani:2000pi}.
At the tree level, the emission of a Higgs boson
from an external leg $i$ leads to a singularity in the soft limit, where the
corresponding top-quark propagator $((p_i+k)^2-m_t^2)^{-1}$
becomes $(p_i^2-m_t^2)^{-1}$, with $p_i^2=m_t^2$. In contrast, the emission of a Higgs boson from an
internal propagator will not lead to any singularity in the soft limit,
since the momentum of the emitting top quark is off-shell.
This fact is not specific to the emission of a soft Higgs boson, and is instead
a well-known feature from the eikonal factorization in QED and QCD.
At the loop level, one has to consider also soft emission from internal
lines that carry a soft momentum in the loop (this includes, for instance,
pentagon loop diagrams), as explained in Ref.~\cite{Catani:2000pi}.
However, in the case of soft Higgs boson radiation,
the internal-line contributions depending on the momenta of two heavy quarks cancel out.
This feature is a
consequence of the fact that the Higgs boson coupling is essentially {\it abelian}
(it commutes with the soft QCD interactions in the loop). Indeed, the same
factorization
of internal line contributions in QCD loop diagrams occurs when considering
soft-photon emission.
Eventually, only diagrams involving a single heavy-quark momentum survive, and their contribution is controlled by the soft-limit of the heavy-quark scalar form factor.

The factorisation formula in Eq.~(2) can also be derived by using the techniques of the low-energy theorems~\cite{Ellis:1975ap,Spira:1995rr,Shifman:1979eb,Kniehl:1995tn}. The basic observation is that at the \textit{bare} level we have
\begin{equation}
\lim_{k \rightarrow 0} {\cal M}^{\rm bare}(\{p_i\},k) = \frac{m_{t,0}}{v}\sum_{i=3,4}\,\frac{m_{t,0}}{p_i \cdot k}\, {\cal M}^{\rm bare}(\{p_i\})\, ,
\end{equation}
where $m_{t,0}$ is the bare mass of the top quark. The renormalisation of the heavy-quark mass and its wave function induces
a modification of the Higgs coupling to the top quark.
The bare amplitude for the emission of a soft Higgs boson from a top quark with momentum $p$ can be written as
\begin{equation}
\lim_{k \rightarrow 0} {\cal M}_{t\to tH}^{\rm bare}(p,k) = \frac{m_{t,0}}{v}\, \frac{\partial}{\partial m_{t,0}}\, {\cal M}_{t\to t}^{\rm bare}(p) \biggl |_{p^2 = m_t^2}\, ,
\label{eq:LET}
\end{equation}
where
\begin{equation}
{\cal M}^{\rm bare}_{t\to t}(p)={\bar t}_0(p)\left(-m_{t,0}-\Sigma(p)\right)t_0(p)\, .
\end{equation}  
By using the results of the ${\cal O}(\as^2)$ contribution to the heavy-quark self-energy $\Sigma(p)$~\cite{Broadhurst:1991fy, Gray:1990yh} we can evaluate
the right-hand side of Eq.~(\ref{eq:LET}).
After renormalising mass and wave function of the heavy quark in the on-shell scheme and the strong coupling in the $\msbar$ scheme,
we derive the effective coupling $F(\as(\muR);m_t/\muR)$ of  Eq.~(\ref{eq:F}), which describes the interaction of the top quark with a soft Higgs boson.

To the best of our knowledge, the factorisation formula in Eq.~(2) has never been presented in the literature beyond tree-level,
and is therefore a new result of this work.
Besides its current application to the construction of an approximation of the virtual $t{\bar t}H$ amplitudes, the factorisation formula in Eq.~(2) can also serve as a non-trivial check of future multi-loop computations. Strictly speaking, the formula holds as \mbox{$k\to 0$}, and, therefore, the limit \mbox{$m_H\ll m_t$} has to be taken as well.
We find that the point-wise difference between the
exact and the approximated matrix elements at tree-level and one-loop order is at the per-mille level for
$E_H<1$ GeV and $m_H=0.5$ GeV. We also note that the factorisation formula in Eq.~(2) can straightforwardly be extended to the production of an arbitrary number of top-quark pairs%
\footnote{We have numerically checked up to one-loop order by using \Recola~\cite{Actis:2016mpe,Denner:2017wsf,Denner:2016kdg} that,
for a very light and soft Higgs boson, the formula holds, besides $\ttbH$, for instance also for $t{\bar t}t{\bar t}H$ production.}.

Given that the soft factorisation formula in Eq.~(2) only describes the leading behaviour
in the $k\to 0$ limit, the radiation of a Higgs boson originated from highly off-shell top propagators
is not captured.
At LO, emission from the final state top quarks is the only Higgs boson production mechanism in the $q{\bar q} \to t{\bar t}H$ partonic channel, whereas in the $gg\to t{\bar t}H$ channel radiation off a virtual top quark contributes as well. This can explain the worse quality of the soft approximation in the $gg$ channel (see Table~I). Starting from NLO, however, diagrams with virtual top quarks radiating a Higgs boson are present in both partonic channels.

We finally note that our approach differs from that used in early NLO calculations~\cite{Dawson:1997im} of $\ttbH$ production and also in the recent study of Ref.~\cite{Brancaccio:2021gcz}.
Our approximation is purely {\it soft}, while in Refs.~\cite{Dawson:1997im,Brancaccio:2021gcz} the Higgs boson is treated as a {\it collinear} parton.

%\twocolumngrid

\bibliography{biblio}

\end{document}